\documentstyle[aps,psfig,multicol]{revtex}
\begin{document}
\newcommand\ds{\displaystyle}
\newcommand\mb[1]{\mathbf{#1}}
\newcommand\be{\begin{equation}}
\newcommand\ee{\end{equation}}
\newcommand\bea{\begin{eqnarray}}
\newcommand\eea{\end{eqnarray}}
\newcommand\ind[1]{{\mbox{\scriptsize\sl #1}}}
\newcommand\ve{\varepsilon}
\newcommand\en{\varepsilon_n}
\newcommand\meff{m_\ind{eff}}
\newcommand\sign{{\mathrm sign}}
\newcommand\vx{v_x}
\newcommand\vxa{\langle v_x\rangle}

\title{Transitions in the horizontal transport of vertically
vibrated granular layers}
\author{Z. Farkas, P. Tegzes, A. Vukics, and T. Vicsek}
\address{Department of Biological Physics, E\"otv\"os University,
Budapest, P\'azm\'any P. Stny 1A, 1117 Hungary}

\maketitle
\begin{abstract}

Motivated by recent advances in the investigation of fluctuation-driven
ratchets and flows in excited granular media, we have carried out
experimental and simulational studies to explore the horizontal
transport of granular particles in a vertically vibrated system whose
base has a sawtooth-shaped profile.  The resulting material flow
exhibits novel collective behavior, both as a function of the number of
layers of particles and the driving frequency; in particular, under
certain conditions, increasing the layer thickness leads to a {\it
reversal of the current}, while the onset of transport as a function of
frequency occurs gradually in a manner
reminiscent of a phase transition.  Our experimental findings
are interpreted here with the help of extensive, event driven Molecular
Dynamics simulations.  In addition to reproducing the experimental
results, the simulations revealed that the current may be {\it reversed}
as a function of the {\it driving frequency} as well.  We also give
details about the simulations so that similar numerical studies can be
carried out in a more straightforward manner in the future.

\end{abstract}


\begin{multicols}{2}

\section{Introduction}

The best known and most common transport mechanisms involve gradients of
external fields or chemical potentials that extend over the distance
traveled by the moving objects.  However, recent theoretical studies
have shown that  there are processes in far from equilibrium systems
possessing vectorial symmetry that can bias thermal noise type
fluctuations and  induce macroscopic motion on the basis of purely
local effects.  This mechanism is expected to be essential for the
operation of molecular combustion motors responsible for many kinds of
biological motion; it has also been demonstrated experimentally in
simple physical systems \cite{RoSaAj94,Fau95}, indicating that it could
lead to new technological developments such as nanoscale devices or
novel types of particle separators.  Motivated by both of these
possibilities, as well as by interesting new results for flows in
excited granular materials
\cite{JaNaBe96,DoFaLa89,ClDuRa92,EhJaKa95,PaBe193,KnJaNa93}, we have
carried out a series of experimental and simulational studies that
explore the manner in which granular particles are {\em horizontally}
transported by means of {\em vertical} vibration.

In the corresponding theoretical models -- known as ``thermal ratchets''
-- fluctuation-driven transport phenomena can be interpreted in terms of
overdamped Brownian particles moving through a periodic but asymmetric,
one-dimensional potential in the presence of nonequilibrium fluctuations
\cite{AjPr92,Mag93,AsBi94,DoHoRi94}.  Typically, a sawtooth-shaped
potential is considered, and the nonlinear fluctuations are represented
either by additional random forces or by switching between two different
potentials.  Collective effects occurring during the fluctuation-driven
motion have also been considered \cite{DeAj96,DeVi95,JuPr95}, leading to
a number of unusual effects that include current reversal as a function
of particle density.  Here we investigate an analogous transport
mechanism for granular materials.  By carrying out experiments -- both
real and numerical -- on granular materials vibrated vertically by a
base with a sawtooth profile, it is possible to achieve a fascinating
combination of two topics of considerable current interest -- ratchets
and granular flows.  A number of recent papers have focused on
vibration-driven granular flow, and the details of the resulting
convection patterns have been examined, both by direct observation
\cite{DoFaLa89,ClDuRa92,PaBe193,KnJaNa93} and by magnetic resonance
imaging \cite{EhJaKa95,NaAlCa93}.  Granular convection has also been
simulated numerically by several groups; the study most closely related
to the present work deals with the horizontal transport that occurs when
the base is forced to vibrate in an asymmetric manner \cite{GaHeSo92}.

\section{Experiments}

We have experimentally investigated of the horizontal flow of granular
material confined between two upright concentric cylinders undergoing
vertical vibration.  In order to induce transport, the height of the
annular base between the cylinders has a periodic, piecewise-linear
profile (in other words, it is sawtooth-like). The experimental setup is
also described in Ref. \cite{DeTeVi}, and is briefly reviewed here.

\subsection{Apparatus}

\begin{figure}
\centerline{\psfig{figure=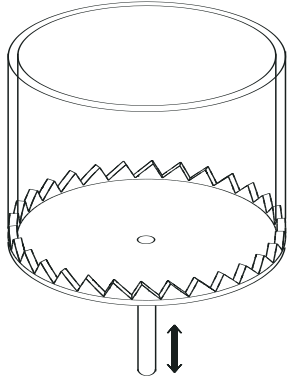,width=0.21\textwidth}}
\caption{The schematic view of the experimental apparatus.}
\label{fig1}
\end{figure}

Figure \ref{fig1}
shows a schematic view of the experimental apparatus.  To
achieve a quasi-two-dimensional system without boundaries in the
direction of the expected flow the granular material is placed between
two concentric glass cylinders \cite{PaBe193}.
The mean diameter of the cylinders is 9{.}2~cm, while the gap between the
cylinders is 3{.}5~mm.
A ring filling the gap between the cylinders, with a sawtooth
profile on its upper surface, is mounted on the base of the container;
the ring is made of either PVC (soft plastic), hard plastic or aluminum,
and different sawtooth shapes are used.  The entire assembly
is vertically vibrated with a displacement that depends sinusoidally on
time.

\begin{figure}
\centerline{\psfig{figure=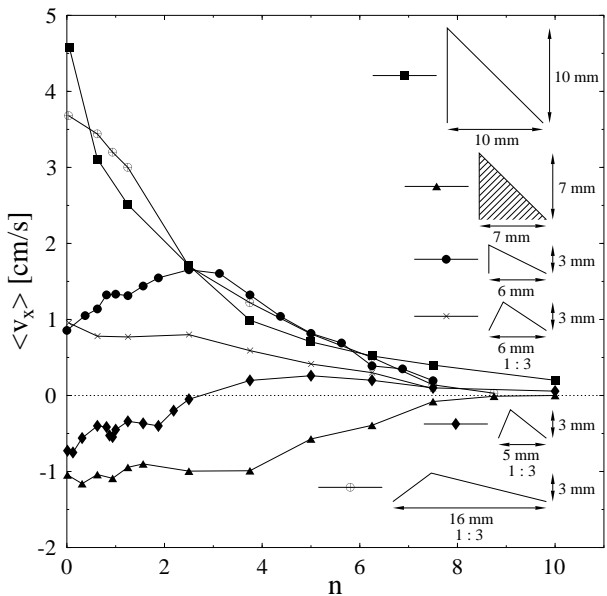,width=0.45\textwidth,bbllx=50pt,bblly=40pt,bburx=590pt,bbury=590pt}}
\caption{The average horizontal velocity of the beads as a function of
height of the granular layer.
Here we characterize the height of the layer by
$n=N/k$, where $N$ is the total number of particles and $k$ is the
number of particles in a single layer.
The various curves represent measurements for various sawtooth shapes
(given by their width $w$, height $h$, and asymmetry parameter $a$,
which is the ratio of the horizontal projection of the left
part of a tooth to its total extension) and
materials. The shape of the sawtooth can be seen in the figure, the
filled sawtooth represents hard plastic material, the others are
made of PVC (soft plastic).  The particles are glass balls.
The amplitude and the frequency are $A=2$~mm, $f=25$~Hz.}
\label{fig2}
\end{figure}

Monodisperse glass balls are used in the experiments,
which are nearly spherical with diameter 3.3 mm $\pm$ 2\%.
As shown in the inset of Fig.\ \ref{fig2},
the size of the sawtooth is in the same range as that of the particles.

\subsection{Results}

Provided the frequency is sufficiently large, the vertical vibration
causes horizontal flow of the entire granular layer. This bulk motion is
reproducible over repeated experiments. The average flow velocity is
determined by tracking individual tracer particles visible through the
transparent cylinder walls. In order to average out fluctuations, the
particles are allowed to travel large distances; depending on the size of
the fluctuations this distance is between 1.5 and 6\,m (equal to 5--20
times the circumference of the system). Each point shown in the graphs is
an average over 3--6 tracer particles.

Figure 2 shows the horizontal flow velocity as a function of the number
of particles for various possible systems.  The actual sawtooth and
particle shapes are also indicated. {\it Positive velocities} are
defined as follows: Moving in the positive direction the first
edge of a sawtooth is the steeper one; i.e., {\it from left to the right} for
these cases.  The vibration amplitude and frequency are $A = 2$\,mm and
$f = 25$\,Hz; the dimensionless acceleration $\Gamma = (2 \pi f)^2 A /
g$ is an important quantity for vibrated granular systems, so that here
$\Gamma = 5$.

We have observed a variety of qualitatively different kinds of behavior,
some of which can be interpreted by simple geometrical arguments.  The
most surprising phenomenon is that in certain cases the velocity changes
sign; in other words, the flow {\em direction} depends on the layer
thickness.  In some cases the curves are monotonically decreasing, while
others have well defined maxima.  Altering the particle shape reduces
the velocity and shifts the location of the maximum, but the shape of
the curve remains unchanged (results not shown here).
Likewise, changing the elasticity of the
base does not alter the curves qualitatively.  The only feature common
to the different curves is that beyond a certain layer thickness the
velocity magnitude decreases as further layers are added.

\begin{figure}
\centerline{\psfig{figure=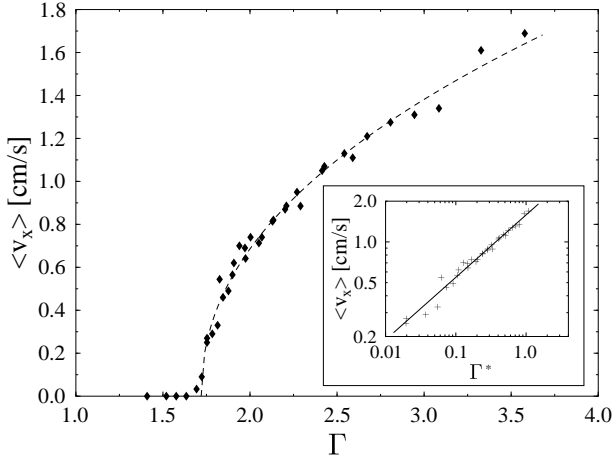,width=0.48\textwidth}}
\caption{Horizontal velocity $\langle v_x\rangle$ as a function
of the dimensionless acceleration $\Gamma$ at constant amplitude
($A=2$~mm). The experiment is for a strongly asymmetric aluminum
sawtooth ($w=12$~mm, $h=7$~mm, $a=0$) and 200 glass balls.
In the inset we display the data on a log-log scale for $\langle v_x\rangle$
close to the transition as a function of $\Gamma^*=(\Gamma-\Gamma_c)/
\Gamma_c$ where $\Gamma_c=1.7$. The slope of the fitted line is $0.48$.}
\label{fig3}
\end{figure}

Figure \ref{fig3} shows the $\Gamma$ dependence of the
flow velocity for a system
of 200 balls (amounting to 4 layers) for constant $A$.
 Flow occurs only
above a critical acceleration $\Gamma_c \simeq 1.7$. Above this critical
value the velocity appears to follow a power law
\begin{equation}
 v(\Gamma) \propto (\Gamma - \Gamma_c)^{0.48},
\end{equation}
suggesting that the onset of flow resembles the kind of phase transition
observed in hydrodynamic instabilities such as thermal convection
\cite{BeDu84}. In this particular case the exponent we obtained is close
to 1/2 indicating that the transition is more like a bifurcation.
However, for other parameters (see, e.g., the simulational results
presented in section \ref{sec:results}) we obtained other exponents as well.

\section{Event driven simulations}

Simulations have been useful in the studies of granular systems
\cite{Bar94,Her95}.
Here we perform event driven simulations of inelastic particles with an
additional shear friction in a two-dimensional system whose
base has a sawtooth-shaped profile.
In event driven simulations
the difficult part of the algorithm is determining the next event
(e.g.,\ the next collision), and
the motion of the particles is calculated
analytically between two events \cite{AllTild87,Wolf96,Lub91}.
The event driven simulations discussed below
represent an alternative to the more common Molecular Dynamics
calculations assuming differentiable interaction potentials between the
particles \cite{Rap95}.

\begin{figure}
\centerline{\psfig{figure=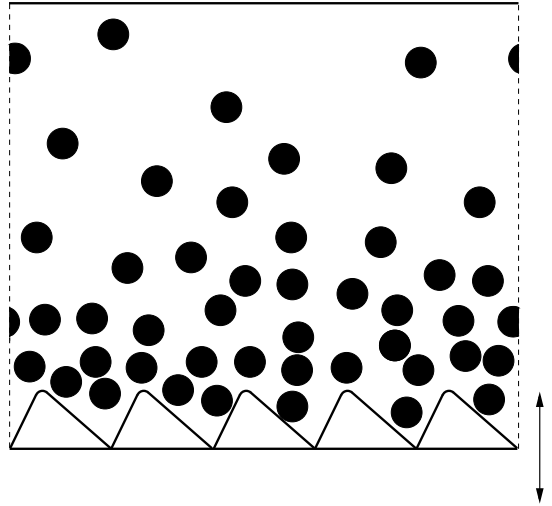,width=0.40\textwidth}}
\caption{Visualization of the arrangement used in the simulations. The
upper wall is fixed,
the sawtooth shaped base is oscillating sinusoidally. Periodic boundary
condition is applied in horizontal direction.}
\label{fig4}
\end{figure}

\begin{figure}
\centerline{\psfig{figure=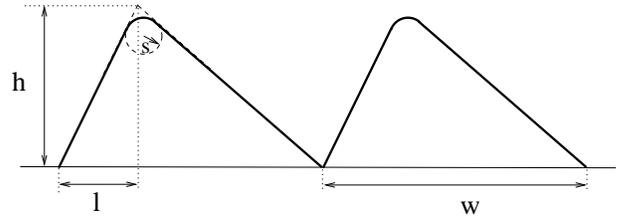,width=0.45\textwidth}}
\caption{The geometry of the sawtooth base.
The shape of a tooth can be described by three parameters: its height $h$,
width  $w$, and asymmetry parameter $a=l/w$, where $l$ is the
horizontal projection of the left part of the tooth.
The top of the teeth are rounded by an arc with radius $s$ smoothly joining
to both sides.}
\label{fig5}
\end{figure}

The particles are modeled as disks with a given radius $r_i$.
These disks can rotate, their moment of inertia is $\ds\frac{1}{2}m_i r_i^2$
($m_i$ is the mass of the $i$th particle).
The motion of the beads is simulated in two-dimensional cell (see Fig.\
\ref{fig4}). To realize the topology of the experiment,
periodical boundary condition is applied in the horizontal direction.
The height of the cell is chosen in such a way that
the number of particles hitting the upper wall is negligible.
The base is oscillating sinusoidally with frequency $f$ and amplitude $A$.
The geometry of the base can be seen in Fig.\ \ref{fig5} in detail.
The shape of a tooth can be described by three parameters:
its height $h$, width $w$ and asymmetry parameter $a$.
This asymmetry parameter is $a=l/w$, where
$l$ is the distance of the projection of the sawtooth's top point
from the left side point of the unit.
The top point of a tooth is rounded by an arc which smoothly joins the
two sides.

We consider collisions during which the particles can stick to or
slide along each other's surface.
For calculating the velocities of two particles after a collision,
a simple model is used which corresponds to recent collision models
\cite{Walton86,Foerster94,Luding95} with $\beta_0=0$, i.e., without
tangential restitution.

The algorithm described in \cite{Lub91}
has been implemented in our simulation with some
extensions, e.g.\ gravitational field, rotating particles, sinusoidally
moving objects.
In the Appendix we give some details of the calculations specific
to this simulation so that similar studies could be carried out or
reproduced in the future in a more direct manner.

A typical problem arising in event driven granular simulations is called
inelastic collapse \cite{McNamYoung92}.
One possibility to avoid the numerical difficulties
associated with it is to make the coefficient of normal restitution
($\varepsilon$) velocity dependent \cite{McNamYoung92,Luding96}.
We use a very simple rule in our simulation:
$\varepsilon$ is constant until $v_n'>v_n^{min}$
($v_n'$ is the relative normal velocity of the colliding particles
after the collision),
however, if $v_n'$ would be less than $v_n^{min}$, it is set to this minimal
normal velocity: $v_n'=v_n^{min}$. Therefore, the relative normal velocity
cannot be less than $v_n^{min}$ between two objects after their collision.
Using $v_n^{min}=5$~mm/s is enough to avoid inelastic collapse, and still
this extra rule does not influence the dynamics of the system,
as far as the transport is concerned. We checked this by determining the
proportion of collisions when this rule had to be used and checking the
change in the average velocity as the threshold value was lowered.
We found that in most situations the
number of collisions when this rule has to be applied is negligible, and
even in dense systems we did not detect a change (beyond error
bars) in the transport velocity when $v_n^{min}$ was decreased.
The simulation program was written in C programming language, and it ran
on personal computers with Linux operating system.

\section{Results}
\label{sec:results}

The system we studied is a complex one, with many parameters.
Since it is impossible to explore the dependence of its behavior
on each of the parameters, some of them were fixed in all
of the simulations. (Still there remained quite a few parameters to vary,
the results shown in this section are selected from simulations with
more than 5000 different parameter sets.)
We fixed two parameters: the width of a sawtooth was $w=6$~mm and
the amplitude of the oscillation was $A=2$~mm.
According to previous results \cite{Fa98}, the {\em direction} of the
horizontal transport
does not depend very much on the friction coefficient $\mu$.
It may, however, depend on the coefficient
of restitution $\varepsilon$, but due to the great number of other
parameters, $\varepsilon$ was kept at fixed value of $0.8$ in all
parameter sets except for one series when it was varied from $0.4$ to $1.0$.
For the sake of simplicity, both $\varepsilon$ and $\mu$ was chosen to be
the same in particle-particle and particle-wall collisions, although
the simulation allows setting different values for different type of
collisions.
The radii of the particles varied from $1.05$~mm to $1.155$~mm uniformly,
to avoid the formation of a hexagonal structure which often appears in
two dimensions when the system freezes.
The masses were equal, hence it did not matter what their actual value was.

\subsection{Dependence on the sawtooth shape}

First we investigated how $\vxa$ depends on the shape of the sawtooth
(other parameters were kept fixed), where the average has been taken
over both time and the individual particles.
We intended to find sawtooth shapes resulting in negative transport.
According to the experiments, negative transport occurs only for a
few particle layers ($n=0\dots 3$), and as the number of
particles is increased, the direction of the transport becomes positive.
We chose $n=1$, and the frequency was $f=25$~Hz ($\Gamma=5.0$).
The width of the cell was $24w=144$~mm, therefore
one layer contained about 70 particles.
These simulations lasted for 50 internal seconds, the following
results represent averages of two runs for each parameter sets
(with different initial conditions).

\begin{figure}
\centerline{\psfig{figure=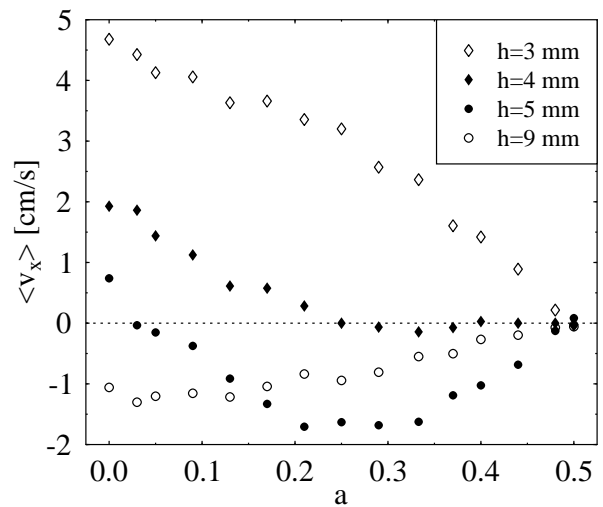,width=0.5\textwidth}}
\caption{The horizontal transport as a function of the asymmetry of
the sawtooth. The shape of the curves depends on the height $h$ of the
sawtooth.}
\label{fig6}
\end{figure}

In Fig.\ \ref{fig6} $\vxa$ can be seen vs.\ $a$ (the
height was fixed at four different values).
The first thing worth to note is that $\vxa$ tends to zero as the shape
becomes symmetrical (i.\ e.\ $a$ tends to $0.5$).
The shape of the plots depends on the height $h$ of the sawtooth.
At $h=3$ mm the velocity is positive and decreases monotonically, while
at $h=5$ mm it is negative and has a minimum at $a\simeq 2.5$, and when
the height is $9$ mm, it is still negative but monotonically increasing.
These curves are helpful in understanding the mechanism of negative
transport.

\begin{figure}
\centerline{\psfig{figure=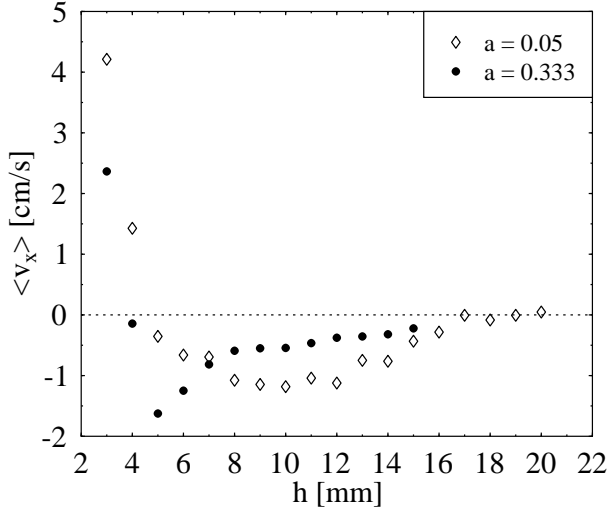,width=0.5\textwidth}}
\caption{Horizontal transport as a funcion of height of the sawtooth.
The shape does not depend very much on the asymmetry of the sawtooth.}
\label{fig7}
\end{figure}

In Fig.\ \ref{fig7} the dependence of the horizontal transport
on the height of the sawtooth can be seen.
The shapes of the graphs are similar in case of $a=0.05$ and $a=0.333$:
$\vxa$ is positive if $h$ is small, and decreases and turns to
negative with increasing $h$. Finally, it becomes zero for high $h$
values.
More detailed investigations showed that it is true for nearly all
possible values of $a$ (except for $a\simeq 0.5$, since then the transport
vanishes).
In conclusion, the horizontal transport can be negative if the height
of the sawtooth is between $4$ and $16$~mm and if the asymmetry
parameter $a$ is less then about $0.4$. In other cases it is positive
or zero (if the sawtooth is symmetric).
It is important to note that these results are valid only for a small
number of particles. As it will be shown, increasing the number
of particles the negative transport vanishes and turns into positive
in all cases.

\begin{figure}
\centerline{\psfig{figure=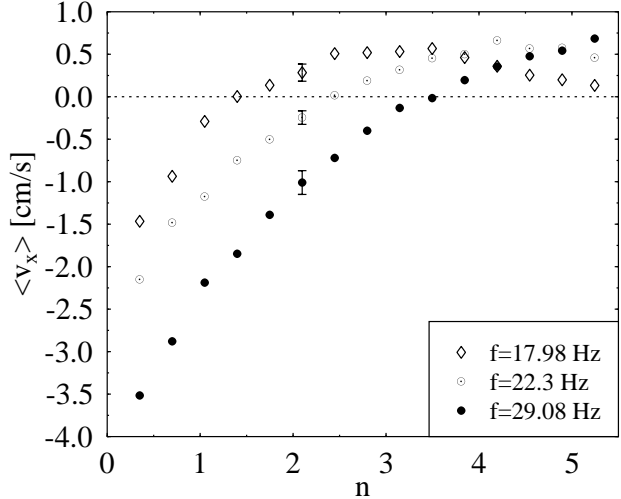,width=0.5\textwidth}}
\caption{The horizontal transport as a function of layer width at three
different frequencies, when the shape of the sawtooth is: $h=6$~mm and
$a=0.25$.
The system width was $10w=60$~mm, one run lasted for 30 internal seconds.
The data is average of 7 runs.}
\label{fig8}
\end{figure}

\subsection{Dependence on the layer width}
The results in the previous section show that if there are only few particles,
then depending on the shape of a  sawtooth, the transport
can be negative or positive. Now we show what happens if more and more
particles are added into the system.

\begin{figure}
\centerline{\psfig{figure=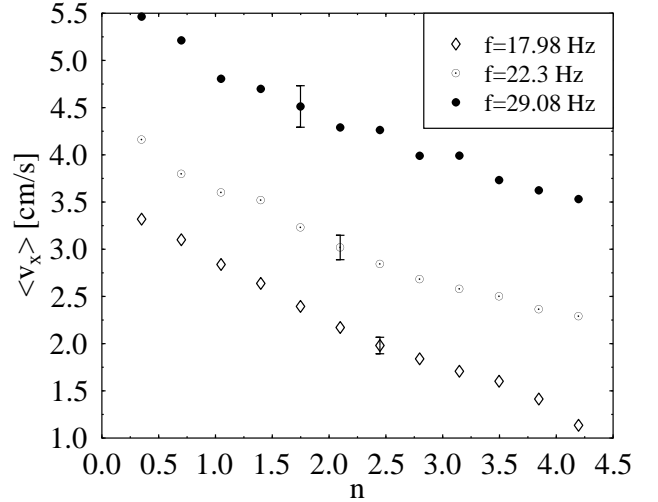,width=0.5\textwidth}}
\caption{The horizontal transport as a function of layer width at three
different frequencies. The shape of the sawtooth: $h=3$~mm and $a=0.1$.
The data shown here is average of 7 runs, one run lasted for 30
internal seconds.}
\label{fig9}
\end{figure}

In Fig.\ \ref{fig8} one can see that negative transport becomes
zero and turns to positive as the layer width is increased.
(Unfortunately
the event-driven simulation becomes unfeasible as the density increases
with increasing particle number, therefore $n\simeq 5$ is the maxium
layer width we can have in our simulation.)
If the sawtooth shape is such that even for a few particles the transport
is positive, then with increasing particle number the transport
decreases (see Fig.\ \ref{fig9}).

\begin{figure}
\centerline{\psfig{figure=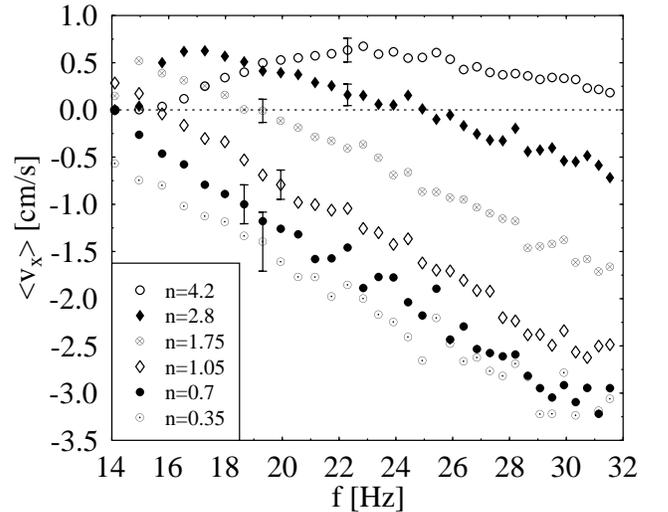,width=0.5\textwidth}}
\caption{Horizontal transport as a function of driving frequency, with
six different layer widths, where the relation between the frequency and
the dimensionless acceleration is $f=(\Gamma g)^{1/2} A^{-1/2} (2\pi)^{-1}$,
with $A=2$~mm.  The shape of the sawtooth: $h=6$~mm, $a=0.25$.
The system width is $10w=60$~mm, one run lasted for 30 internal seconds.
The plots for $n=0.35$, $n=0.7$, $n=1.05$, $n=1.75$, $n=2.8$, and
$n=4.2$ are average of 30, 18, 12, 7, 7, 7 runs, respectively.}
\label{fig10}
\end{figure}

\subsection{Frequency dependence}
Another interesting question is what happens if the frequency is varied.
A sawtooth shape producing negative transport for few particles was
chosen.
Then with different layer widths the horizontal transport shows interesting
behavior as a function of frequency. The results are shown in Fig.\
\ref{fig10}.

Generally, the negative transport becomes stronger with increasing
frequency. In three cases ($n=1.05$, $n=1.75$, and $n=2.8$) the transport
{\em reverses}, from positive to negative.
This phenomenon is again very illuminating when we try to explain the
mechanism of negative transport.

\subsection{Dependence on the coefficient of restitution}

\begin{figure}
\centerline{\psfig{figure=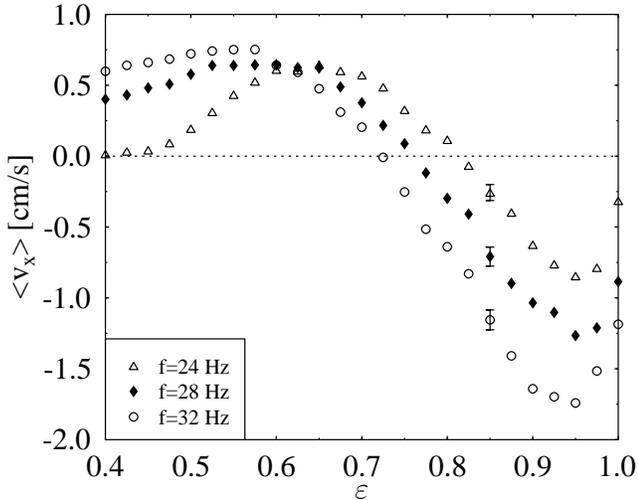,width=0.5\textwidth}}
\caption{The horizontal transport as a function of
coefficient of restitution for three different driving freqencies.
The layer width is $n=2.8$, the shape of the sawtooth and the system
width are the same as in Fig.\ \ref{fig10}. The data shown here are
averaged over 20 runs.}
\label{fig11}
\end{figure}

We have demonstrated that the transport properties of the granular layer
in our system depend on the various parameters in a non-trivial way.  In
particular, increasing the layer width, through its effect on the
enhanced clustering of the beads, resulted in a change of the direction
of the current.  Thus, we are motivated to investigate the effect of
changing the coefficient of restitution as well.  Except for these
series, $\varepsilon$ is fixed at $0.8$, but now it varies from $0.4$ to
$1.0$.  It can be seen in Fig.\ \ref{fig11} that as the dissipation is
decreased the direction of transport reverses from positive to negative.
The magnitude of the velocity reaches a local maximum near
$\varepsilon=0.95$, and decreases until $1.0$.

\begin{figure}
\centerline{\psfig{figure=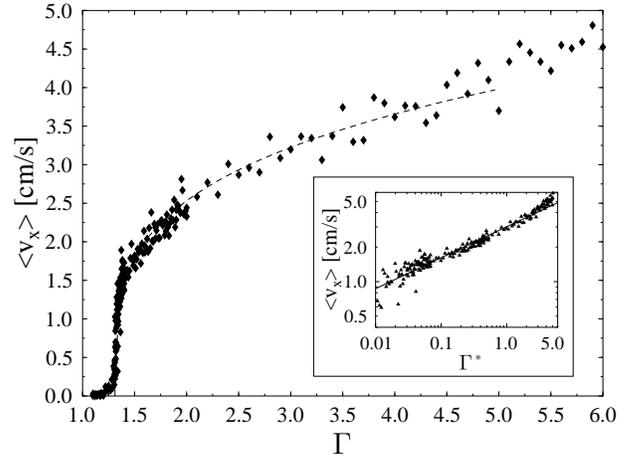,width=0.44\textwidth,bbllx=40pt,bblly=15pt,bburx=585pt,bbury=445pt}}
\caption{Horizontal velocity $\vxa$ as a function of $\Gamma$.
The shape of a sawtooth: $h=3$~mm, $a=0.1$. The layer width was
$n=0.7$, the system width $10w=60$~mm. The simulations ran for 30 internal
seconds. A power low function was fitted:
$\Gamma^{*\gamma}$ with $\Gamma^*=(\Gamma-\Gamma_c)/\Gamma_c$.
The fitted values are $\Gamma_c=1.31\pm 0.03$, and $\gamma=0.27\pm 0.05$.}
\label{fig12}
\end{figure}

\begin{figure}
\centerline{\psfig{figure=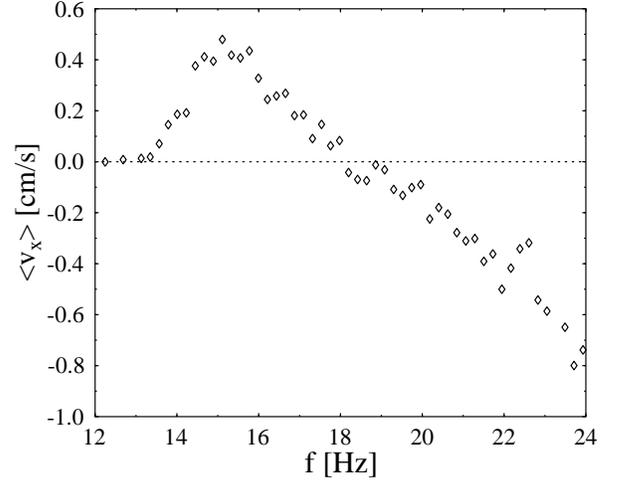,width=0.48\textwidth}}
\caption{Horizontal transport velocity as a funcion of driving frequency.
The shape of a sawtooth: $h=6$~mm, $a=0.25$. The layer width is $n=1.75$.
The system width is $10w=60$~mm, the simulations ran for 30 internal seconds.}
\label{fig13}
\end{figure}

\subsection{Phase transition}
When the driving freqency is not high enough (i.e., the dimensionless
acceleration $\Gamma<1$), then  $\vxa=0$, since
the whole granular layer has zero velocity compared with the oscillating
base (solid phase).
However, the transport may remain zero for $\Gamma>1$, until a critical
acceleration. Increasing the frequency, some  of the particles, and later
all of them are flying (fluidized phase), therefore, transport may
appear.
We investigated this phase transition (with control parameter $\vxa$)
in case of two different sawtooth shapes.
First we chose a shape which produces positive transport for a few
particles.
Figure \ref{fig12} shows the $\Gamma$ dependence of the velocity.
The transport appears when $\Gamma>\Gamma_c=1.31$. Above this, a power
law of the form $\vxa\sim (\Gamma-\Gamma_c)^{0.27}$ can be fitted to
the results.

Choosing a sawtooth for which the transport velocity is negative for
small particle number, one expects different behavior.
Fig.\ \ref{fig13} shows how the velocity depends on the frequency
in such a case. The velocity becomes positive growing linearly above the
critical
frequency, and increasing the frequency, it turns back, and the transport
reverses.
\end{multicols}
\begin{figure}
\centerline{\psfig{figure=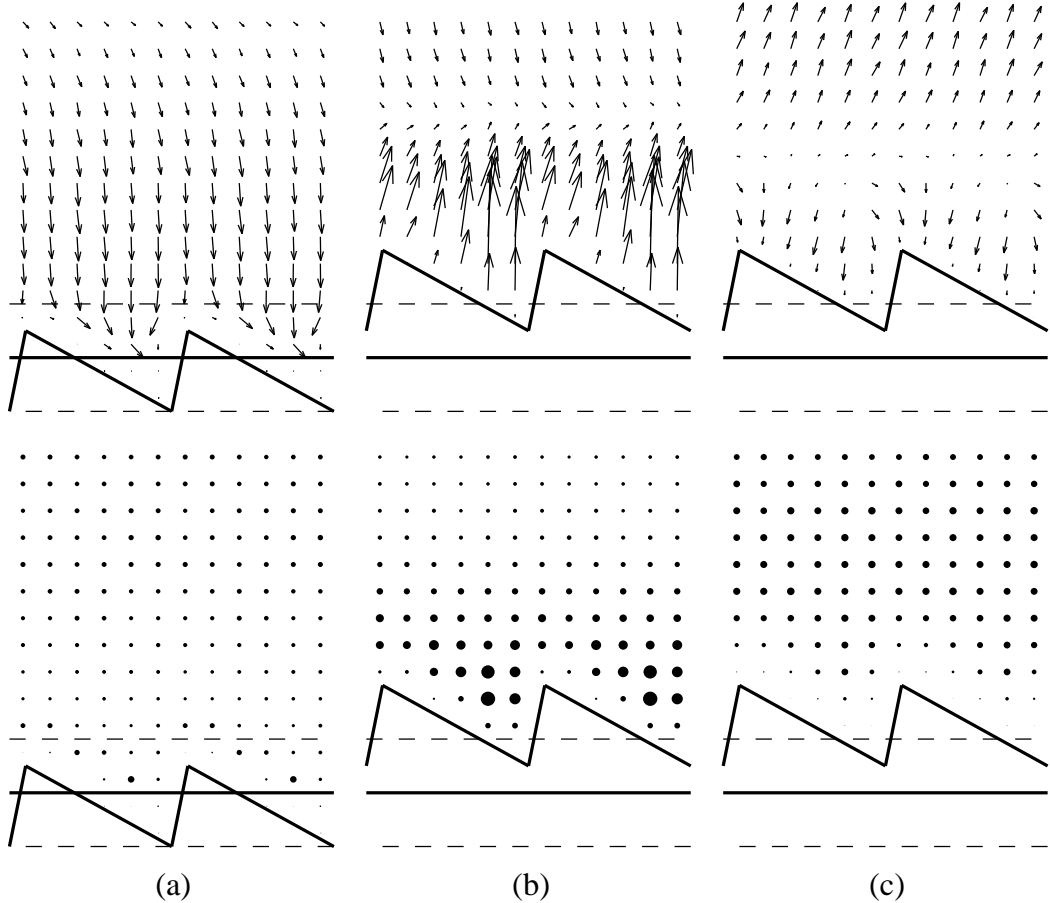,width=0.77\textwidth}}
\caption{Momentum density field (upper row) and mass density field (lower row)
in case of {\em positive} horizontal transport
($\vxa=3.9\pm0.3$~cm/s, $h=3$~mm, $a=0.1$, $f=25$~Hz, $n=0.7$).
The arrows represent the average momentum vector in a box (with width $w/6$)
starting from the center of the box,
the radii of the disks are proportional to the average mass.
(The momentum and mass carried by a particle whose center is in the box
are taken into account when calculating the average in the box.)
The momentum and the mass are averaged for these spatial boxes and
phase frames of the oscillation (with length $2\pi/6$).
In the $j$th frame, the phase $\phi=2\pi ft$ of oscillation $A\sin(2\pi ft)$
is $j 2\pi/6\le \phi < (j+1)2\pi/6$.
(a) $j=4$, (b) $j=0$, (c) $j=2$. The actual height of the simulation cell
is three times larger than shown here.}
\label{fig14}
\end{figure}

\begin{multicols}{2}

\section{Discussion}

According to our studies of a simplified geometrical model the following
qualitative argument can be used to explain the observed current
reversal as a function of the particle number: There is an intermediate
size and asymmetry of the teeth for
which a single ball falling from a range of near-vertical angles bounces
back to the left (negative direction) in most of the cases.  This effect
is enhanced by rotation, due to friction between the ball and the tooth.
However, if there are many particles present, this mechanism is
destroyed, and on average, the direction of the motion of particles will
become positive (the ``natural'' direction for this geometry); this
corresponds to the usual ratcheting mechanism characterized by larger
distances traveled by the particles along the smaller slope with
occasional jumps over to the next valley between the teeth
\cite{AjPr92,Mag93,AsBi94,DoHoRi94,Bug87}.  There is no net current for
symmetric teeth in our case, however the motion of a single particle
is very interesting in that situation as well \cite{Duran}.

The reversal of the current as a function of the frequency can be
interpreted in a similar manner. For smaller frequencies there are many
particles close to the base and the current is positive. For larger
frequencies the granular state is highly fluidized, the density
considerably decays and the mechanism leading to negative transport for
a small number of particles (discussed above) comes into play.

\begin{figure}
\centerline{\psfig{figure=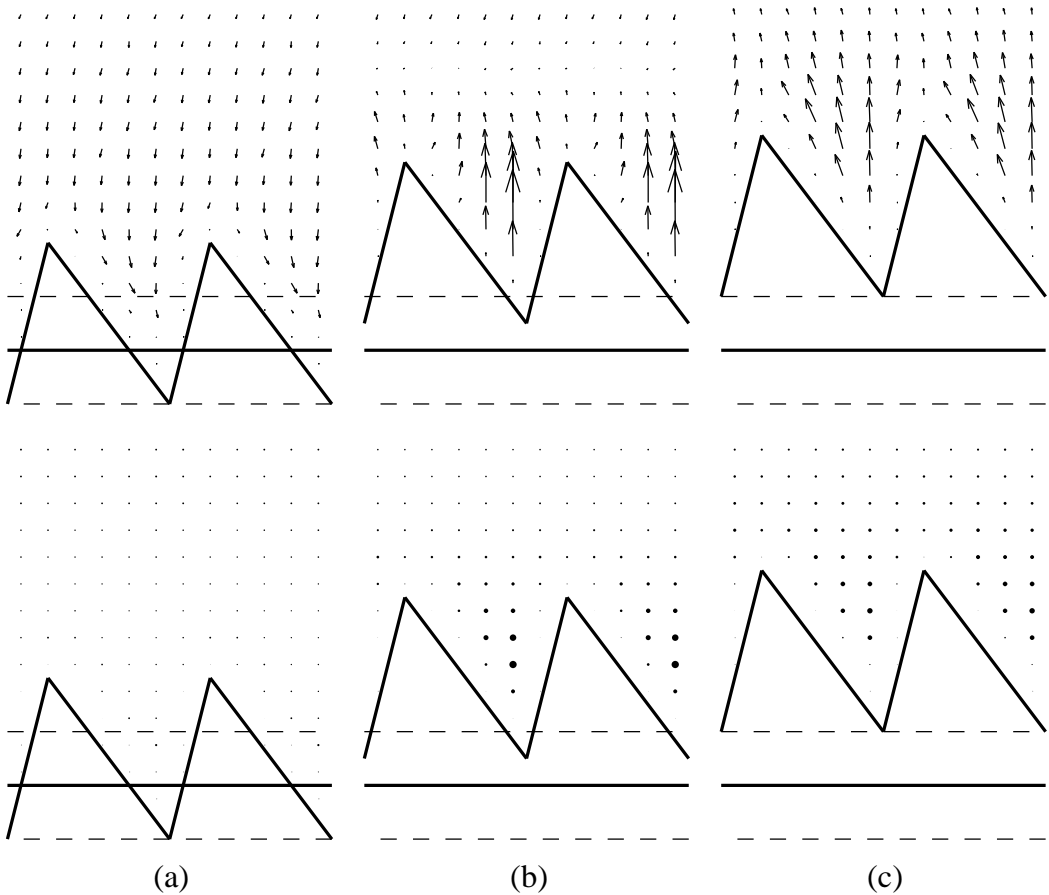,width=0.77\textwidth}}
\caption{The same as in Fig.\ \ref{fig14}, but in case of
{\em negative} horizontal transport ($\vxa=-3.2\pm0.3$~cm/s, $h=6$~mm,
$a=0.25$, $f=30$~Hz, $n=0.35$). (a) $j=4$, (b) $j=0$, (c) $j=1$.
The ratio of the absolute value of the average momentum in a box and
the length of the arrow is the same as in Fig. \ref{fig14}.}
\label{fig15}
\end{figure}

To obtain a deeper insight into the process of horizontal flow,
the momentum density field and mass distribution is plotted in case of
{\em positive}
(Fig.\ \ref{fig14}) and {\em negative}
(Fig.\ \ref{fig15}) transport.
In Fig.\ \ref{fig14} it can be seen that the particles
slide down the side with the smaller slope, and when the base is in
rising phase, they are given a positive net horizontal impulse.
It is worth noting that most particles do not collide with the opposite
side. On the other hand, if the sawtooth is higher and more symmetric,
the off-bouncing particles collide with the opposite side of the sawtooth,
therefore their horizontal impulse is reveresed, and the net current
becomes negative (see Fig.\ \ref{fig14}).
\end{multicols}

\begin{multicols}{2}
We interpret the existence of maxima in Fig.\ \ref{fig2} as a
function of the particle number as follows: If only a few
particles are are present, their motion is erratic, with large jumps
in random directions.  As the number of particles is increased, due to a
inelastic collapse-like process, the particles start to move coherently,
and this more ordered motion, together with the right
frequency, seems to give rise to a kind of resonant behavior as far as
transport is concerned.  Similar maxima can also be observed in
molecular motor calculations and simulations.  Thick layers move
more slowly because of inelastic damping.

The remarkable result that emerges from both experiment and simulation is
that the flow direction can change as the layer thickness varies. This is
entirely unexpected and requires further investigation; the only related
behavior of which we are aware is the alternating current direction in a
model of collectively moving interacting Brownian particles in a
``flashing'' ratchet potential \cite{DeAj96}.


In conclusion, we have investigated granular transport in a system
inspired by models of molecular motors and have observed, both
experimentally and numerically, that the behavior depends in a complex
manner on the parameters characterizing the system. These results ought to
stimulate further research into this fascinating class of problems.

\section*{Acknowledgments}
Useful discussions with I. Der\'enyi are acknowledged.
One of the authors (T. V.) has had extensive interactions with D. Rapaport
about the standard (not event driven) simulations of the present system
(to be published).  This work was supported in part by the Hungarian
Research Foundation Grant No.\ T019299 and MKM FKFP Grant No. 0203/1997.

\appendix
\section*{Collisions as events}

In an event driven algorithm the pace of the simulation is determined
by the time sequences between collisions.
In our simulation, which uses Lubachevsky's event driven algorithm
\cite{Lub91} with extensions, there are four types of collisions,
i.e., a particle may collide with a (i) particle, (ii) resting segment,
(iii) oscillating segment, and (iv) oscillating arc.
To find the next collision is the heart of the matter.
One has to calculate the times of the possible
collisions and find the closest one.
Let us denote the time of the last event by $t_{LE}$.
Now we show
the calculation of the collision time in full detail in all cases.

\subsection{Particle--particle collision}

The motion of particle $i$ until its next collision in a gravitational
field is determined by its position ${\mb r}_{0i}$ and velocity ${\mb v}_{0i}$
at time $t_i$. As the two particles ($1$ and $2$) are moving along
parabolic paths, their position at time $t$ are
\bea
{\mb r}_1(t) &=& {\mb r}_{01}+{\mb v}_{01}(t-t_1)-{\mb j}\frac{g}{2}(t-t_1)^2
\label{r_1_p-p}\\
{\mb r}_2(t) &=& {\mb r}_{02}+{\mb v}_{02}(t-t_2)-{\mb j}\frac{g}{2}(t-t_2)^2,
\label{r_2_p-p}
\eea
where ${\mb j}$ is a unit vector having opposite direction to that of
gravity. A collision occurs if $|{\mb r}_1(t)-{\mb r}_2(t)|=R_1+R_2$,
which can be written, using (\ref{r_1_p-p}-\ref{r_2_p-p}) and
introducing the following notations
\bea
{\mb p} &=& {\mb v}_{01}-{\mb v}_{02}+{\mb j} g (t_1-t_2)\nonumber\\
{\mb q} &=& {\mb r}_{01}-{\mb r}_{02}+{\mb v}_{02}t_2-{\mb v}_{01}t_1-
{\mb j}\frac{g}{2}(t_1^2-t_2^2)\nonumber
\eea
as
\be
|{\mb p}t+{\mb q}|=R_1+R_2.\label{cond_p-p}
\ee
Taking the square of (\ref{cond_p-p}) and divide it by $2$ we get
\be
\frac{{\mb p}^2}{2}t^2+{\mb p\cdot q}t+\frac{{\mb q}^2-(R_1+R_2)^2}{2}=0.
\ee
The two particles can collide with each other only if $|{\mb p}|\ne 0$ and
\be
D=({\mb p\cdot q})^2-{\mb p}^2({\mb q}^2-(R_1+R_2)^2)
\ee
\begin{figure}
\centerline{\psfig{figure=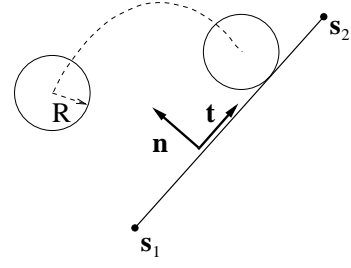,width=0.25\textwidth}}
\caption{Particle colliding with a resting segment. The particle arrives
from the inside of the simulation space and collides with the segment
between its ending points ${\mb s}_1$ and ${\mb s}_2$.}
\label{fig16}
\end{figure}
\par\noindent
is nonnegative, otherwise the particles
will not collide.
If $D>0$, then the two times when the condition of touching is
fulfilled are
\be
t^{coll}_{1,2} = \frac{-{\mb p\cdot q}\pm\sqrt{D}}{{\mb p}^2},
\ee
where $t^{coll}_1 < t^{coll}_2$.
(If $D=0$, then $t^{coll}_1=t^{coll}_2$, and this means that the two
particles only touch each other, so we may consider that no collision
occurs in this case.) Since the smaller solution gives the time of
the first touching, and only collisions in the future should be regarded,
the actual time of the collision is given by $t^{coll}_1$ if $t^{coll}_1>
t_{LE}$, otherwise no collision occurs.
Summarizing the results, two particles collide, and then the collision time
is $t^{coll}_1$, if and only if all of the following three conditions
are fulfilled: $|{\mb p}|\ne 0$, $D>0$, and $t^{coll}_1>t_{LE}$.

\subsection{Particle--resting segment collision}

 Let us denote the position and the
velocity of the particle by ${\mb r}_0$ and ${\mb v}_0$ respectively
at time $t_0$. Let ${\mb n}$ be a unit vector perpendicular to the
segment, pointing towards the inside of the simulation space, and
${\mb s}_1$ and ${\mb s}_2$ the left and right side end point of the
segment with respect to ${\mb n}$ (see Fig.\ \ref{fig16}).
Since the particle is moving along a parabolic path in the gravitational
field, its position as a function of time is
\be
{\mb r}(t)={\mb r}_0+{\mb v}_0 (t-t_0)-{\mb j}\frac{g}{2}(t-t_0)^2,
\label{eq:part-line-pos}
\ee
while its speed is
\be
{\mb v}(t)={\mb v}_0-{\mb j}g(t-t_0).
\ee
The particle collides with the line if
\be
({\mb r}(t)-{\mb s}_1)\cdot{\mb n}=R,
\label{eq:part-line-cond}
\ee
which, after inserting ${\mb r}(t)$ from (\ref{eq:part-line-pos}) and
using the following notations:
\bea
u=-\frac{g}{2}{\mb j\cdot n},\;\;
v=({\mb j}gt_0+{\mb v}_0)\cdot{\mb n},\nonumber\\
w=({\mb r}_0-{\mb s}_1-{\mb v}_0 t_0-{\mb j}\frac{g}{2} t_0^2)\cdot
{\mb n}-R,\;\;
D=v^2-4uv\nonumber
\eea
can be written as
\be
ut^2+vt+w=0.
\label{eq:part-rest-line-cond}
\ee
We have to consider several cases. First, if $u=0$ (in other
words ${\mb j}$ and ${\mb n}$ are perpendicular to each other, therefore
the segment is vertical), then at most one solution is possible.
If $v=0$ in this case, we can say that no collision occurs. If $v\ne 0$,
the solution is
\be
t^{coll}_0=-\frac{w}{v},
\ee
and this is valid only if $t^{coll}_0 \ge t_{LE}$. If $u\ne 0$, then, provided
that $D>0$, there are different solutions:
\be
t^{coll}_{1,2} = \frac{-v\pm\sqrt{D}}{2u}.
\ee
If $D\le 0$, then we have either one (the particle just touches the line)
or no solution. In these cases no collision occurs. Now let us assume that
(\ref{eq:part-rest-line-cond}) has one or more solutions, which can be
$t^{coll}_0$, $t^{coll}_1$, or $t^{coll}_2$. Nevertheless, two more
conditions have to be fulfilled at the same time:
\begin{itemize}
\item[(i)] at the moment of collision the particle has to come from the
inside of the simulation space, i.e.,
\be
{\mb v}(t^{coll})\cdot{\mb n} < 0,
\label{eq:part-rest-line-cond2}
\ee
where $t^{coll}$ denotes a solution. It is clear that even if there are
two solutions, after (\ref{eq:part-rest-line-cond2}) only one remains.
\item[(ii)] It is not enough to collide with the line of the segment,
the particle has to collide with
it between its ending points ${\mb s}_1$ and ${\mb s}_2$.
This is true if
\be
({\mb r}(t^{coll})-{\mb s}_1)\cdot{\mb t}\ge 0\;\;\; \mbox{and}\;\;\;
({\mb r}(t^{coll})-{\mb s}_2)\cdot{\mb t}\le 0,
\ee
where ${\mb t}=\frac{{\mb s}_2-{\mb s}_1}{|{\mb s}_2-{\mb s}_1|}$, a unit
vector parallel to the segment.
Summarizing, a particle collides with a segment only if
(\ref{eq:part-rest-line-cond}) has solution, and with this solution both
(i) and (ii) conditions are fulfilled.

\end{itemize}

\subsection{Particle--oscillating segment collision}
In this case the time of collision cannot be obtained in a closed form,
a numerical method has to be used.
The segment (and so its left ending point) is oscillating in
vertical direction:
\begin{figure}
\centerline{\psfig{figure=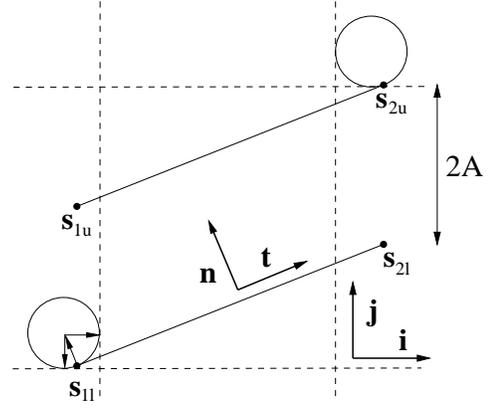,width=0.35\textwidth}}
\caption{A particle can collide with an oscillating segment only if
it enters the area bounded by the vertical dashed
lines and its lowest point is between the two horizontal dashed lines.}
\label{fig17}
\end{figure}
\be
{\mb s}_1(t) = {\mb s}_{10} + {\mb j}A\sin(2\pi ft+\phi_0)
\label{eq:oscillating-line}
\ee
Let the particle has the same properties as in the previous (particle-resting
segment collision) case. Then we get the equation for collision by
replacing ${\mb s}_1$ by ${\mb s}_1(t)$ in (\ref{eq:part-line-cond}).
We can assume that ${\mb jn}\ne 0$, so the segment is not vertical (if
it is, then with a slight modification it can be taken as a resting
segment). Using the notations
\bea
u = -\frac{g}{2A},\;\; b=\frac{({\mb j}gt_0+{\mb v}_0)\cdot{\mb n}}
{A{\mb j\cdot n}},
\nonumber\\
w=\frac{(2{\mb r}_0-2{\mb s}_{10}-2{\mb v}_0t_0-{\mb j}gt_0^2)\cdot{\mb n}}
{2A{\mb j\cdot n}},\nonumber
\eea
the equation is
\be
ut^2+vt+w = A\sin(2\pi ft+\phi_0).
\label{eq:part-osc-line-cond}
\ee
This equation can be solved only numerically. First, we have to find
the time region in which a solution is possible.
In Fig.\ \ref{fig17} one can see the segment in the lowest
and the highest positions, and two particles touching the segment at its
left end point ${\mb s}_{1l}$ (in the lowest postion) and right end point
${\mb s}_{2u}$ (in the hightest position).
It is easy to see that the particle can collide with
the segment if (i) it
enters the area bounded by the two vertical dashed lines and
(ii) the particle's lowest point is in the area bounded by the
two horizontal dashed lines. Regarding these dashed lines as resting segments,
using our previous results, we can determine the time intervals $T_v$ and
$T_h$ when the particle is in a position that conditions (i) and (ii) are
fulfilled. For this, one point of each of these segments
has to be determined (as their direction is known).
The center of the particle touching the
segment at point ${\mb s}_{1l}$ is at point ${\mb s}_{1l}+R{\mb n}$,
so one point of the vertical boundary line is ${\mb s}_{1l}+R{\mb n}+
R({\mb j\cdot n}){\mb i}$, and that of the horizontal boundary line is
${\mb s}_{1l}+R{\mb n}-R({\mb j\cdot n}){\mb j}$. Similarily, for point
${\mb s}_{2u}$, these are ${\mb s}_{2u}+R{\mb n}-R({\mb j\cdot n}){\mb i}$ and
${\mb s}_{2u}+R{\mb n}-R({\mb j\cdot n}){\mb j}$, respectively.
The dot product ${\mb jn}$ is needed to ensure the validity of our results
if ${\mb n}$ is pointing downwards. (In this case `left' and `right', and
`up' and `down' also have to be swapped.)
The time interval $T_v$, when the particle is between the
vertical dashed lines, is a closed interval (in a special case, when the
particle is moving vertically, this can include all time).
The time interval $T_h$, while the lowest point of the particle is between
the horizontal dashed lines, can be empty or one closed interval or
two disjunctive closed intervals. Furthermore, only solutions in the future
are valid, i.e., a solution has to fall into the time interval
$T_*=[t_{LE},\infty[$.
Summarizing, a particle collides with a segment if (\ref{eq:part-osc-line-cond})
has solution in the time interval $T=T_h\cap T_v\cap T_*$, and if there
are more than one of them, the smallest one is the time of collision.
In our simulation the {\em bisection method} \cite{numrec} was used,
with ending tolerance $10^{-12}$~s.

\begin{figure}
\centerline{\psfig{figure=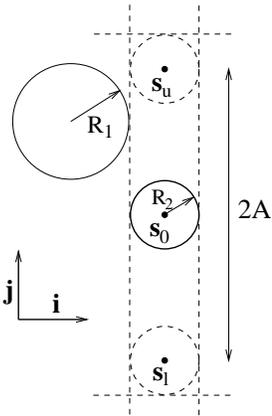,width=0.2\textwidth}}
\caption{A particle can collide with an oscillating circle only if
it enters the rectangular area bounded by the vertical and horizontal
dashed lines.}
\label{fig18}
\end{figure}
\subsection{Particle--oscillating circle collision}

The method of finding the collision time is similar to that of in the
previous case. The center of the circle is moving according to
\be
{\mb s}(t)={\mb s}_0+{\mb j}A\sin(2\pi f t+\phi_0)
\ee
(see Fig.\ \ref{fig18}).
The particle with radius $R_1$
is moving again along a
parabolic path, its position is given by (\ref{eq:part-line-pos}).
The equation for the collision time is
\be
|{\mb r}(t)-{\mb s}(t)|=R_1+R_2,
\label{part-osc-circ}
\ee
where $R_2$ is the radius of the circle.
It is a transcendental equation, and we use again the bisection method for
solving it. The collision may occur only if the the particle enters
the rectangular area bounded by the two horizontal and vertical lines
in Fig.\ \ref{fig18}. It is easy to determine these lines:
the horizontal lines contain points ${\mb s}_l-R_2{\mb j}$ and
${\mb s}_u+R_2{\mb j}$, while the vertical lines contain points
${\mb s}_0\pm R_2 {\mb i}$.
The time region of possible collision can be calculated similarily to the
case of particle--oscillating segment collision.

\end{multicols}

\begin{references}

\bibitem{RoSaAj94} J. Rousselet, L. Salome, A. Ajdari, and J. Prost,
Nature {\bf 370}, 446 (1994).

\bibitem{Fau95} L. P. Faucheux, L. S. Bourdieu, P. D. Kaplan, and A. J.
Libchaber, \prl {\bf 74}, 1504 (1995).

\bibitem{JaNaBe96} H. M. Jaeger, S. R. Nagel, and R. P. Behringer,
Rev. Mod. Phys.,  {\bf 68}, 1259 (1996).

\bibitem{DoFaLa89} S. Douady, S. Fauve, and C. Laroche, Europhys. Lett.
{\bf 8}, 621 (1989).

\bibitem{ClDuRa92} E. Cl\'ement, J. Duran, and J. Rachjenbach, \prl {\bf
69}, 1189 (1992).

\bibitem{EhJaKa95} E. E. Ehrichs, H. M. Jaeger, G. S. Karczmar, J. B.
Knight, V. Yu. Kuperman, and S. R. Nagel, Science {\bf 267}, 1632 (1995).

\bibitem{PaBe193} H. K. Pak and R. P. Behringer, \prl {\bf 71}, 1832
(1993).

\bibitem{KnJaNa93} J. B. Knight, H. M. Jaeger, and S. R. Nagel, \prl {\bf
70}, 3728 (1993).

\bibitem{AjPr92} A. Ajdari and J. Prost, C. R. Acad. Sci. Paris {\bf 315},
1635 (1992).

\bibitem{Mag93} M. O. Magnasco, \prl {\bf 71}, 1477 (1993).

\bibitem{AsBi94} R. D. Astumian and M. Bier, \prl {\bf 72}, 1766 (1994)

\bibitem{DoHoRi94} C. R. Doering, W. Horsthemke, and J. Riordan, \prl {\bf
72},  2984 (1994).

\bibitem{DeAj96} I. Der\'enyi and A. Ajdari, \pre {\bf 54}, R5 (1996).

\bibitem{DeVi95} I. Der\'enyi and T. Vicsek, \prl {\bf 75}, 374 (1995).

\bibitem{JuPr95} F. J\"ulicher and J. Prost, \prl {\bf 75}, 2618 (1995).

\bibitem{NaAlCa93} M. Nakagawa, S. A. Altobelli, A. Caprhan, E. Fukushima,
and E-K. Jeong, Exp. Fluids {\bf 16}, 54 (1993).

\bibitem{GaHeSo92} J. A. C. Gallas, H. J. Herrmann, and S. Sokolowski, J.
Phys. (France) II {\bf 2}, 1389 (1992).

\bibitem{DeTeVi} I. Der\'enyi, P. Tegzes, and T. Vicsek {\it Chaos} {\bf
8} 657 (1998).

\bibitem{BeDu84} P. Berg\'e and M. Dubois, Contemp. Phys. {\bf 25}, 535
(1984).

\bibitem{Bar94} G. C. Barker, in {\em Granular Matter: An
Interdisciplinary Approach}, edited by A. Mehta (Springer, Heidelberg,
1994), p.\ 35.

\bibitem{Her95} H. J. Herrmann, in {\em 3rd Granada Lectures in
Computational Physics}, edited by P. L. Garrido and J. Marro (Springer,
Heidelberg, 1995), p.\ 67.


\bibitem{Duran} J. Duran, Europhys.  Lett., {\bf 17}, 679
(1992)

\bibitem{Bug87} A. L. R. Bug and B. J. Berne \prl {\bf 59},
948 (1987)

\bibitem{AllTild87} M. P. Allen and D. J. Tildesley,
{\em Computer Simulation of Liquids} (Clarendon Press, Oxford, 1987)

\bibitem{Wolf96} D. E. Wolf, in {\em Computational Physics: Selected
Methods -- Simple Exercises -- Serious Applications}, edited by
K. H. Hoffmann and M. Schreiber (Springer, Heidelberg, 1996), p.\ 64.

\bibitem{Lub91} B. D. Lubachevsky, J. Comput. Phys. {\bf 94}, 255 (1991)

\bibitem{Rap95} D. C. Rapaport, {\em The Art of Molecular Dynamics
Simulation} (Cambridge University Press, Cambridge, 1995).

\bibitem{Walton86} O. R. Walton and R. L. Braun, J. Rheol.\ {\bf 30},
949 (1986)

\bibitem{Foerster94} S. F. Foerster, M. Y. Louge, H. Chang, and K. Allia,
Phys.\ Fluids {\bf 6}, 1108 (1994)

\bibitem{Luding95} S. Luding, \pre {\bf 52}, 4442 (1995)

\bibitem{McNamYoung92} S. McNamara and W. R. Young, Phys.\ Fluids A
{\bf 4}, 496 (1992).

\bibitem{Luding96} S. Luding, E. Cl\'ement, J. Rajchenbach, and
J. Duran, Europhys.\ Lett.\ {\bf 36}, 247 (1996).

\bibitem{Fa98} Z. Farkas, Diploma thesis, E\"otv\"os University, 1998.

\bibitem{numrec} W.\ H.\ Press, S. A. Teukolsky, W. T. Vetterling, and
B. P. Flannery, {\em Numerical Recepies in C}
(Cambridge University Press, Cambridge, 1992).

\end{references}
\end{document}